\documentclass[10pt,a4paper,preprint]{elsarticle}
\usepackage{amsmath,amssymb}
\usepackage{graphicx}
\biboptions{sort&compress}

\begin{document}
\begin{frontmatter}
\title{Generalized seniority isomers in and around $Z=82$ closed shell: a survey of Hg, Pb and Po isotopes}
\author{Bhoomika Maheshwari\corref{mycorrespondingauthor}}
\cortext[mycorrespondingauthor]{Corresponding author}
\ead{bhoomika.physics@gmail.com}
\address{Amity Institute of Nuclear Science and Technology, Amity University UP, 201313 Noida, India.}
\author{Deepika Choudhury}
\address{Department of Physics, Indian Institute of Technology Ropar, 140001 Rupnagar, India.} 
\author{Ashok Kumar Jain}
\address{Amity Institute of Nuclear Science and Technology, Amity University UP, 201313 Noida, India.}
\begin{abstract}
  In this paper, we investigate the generalized seniority scheme and the validity of Generalized Seniority Schmidt Model in and around the $Z=82$ semi-magic region. A consistently same multi-j configuration is used to explain all the nuclear spectroscopic properties such as $g-$factors, $Q-$moments and $B(E2)$ trends for the ${13/2}^+$, ${12}^+$ and ${33/2}^+$ isomers in all the three Hg, Pb and Po isotopic chains. The inverted parabolic $B(E2)$ trends for the first $2^+$ states in Hg, Pb and Po isotopes are also explained using the generalized seniority scheme. A comparison with the experimental data is presented, wherever possible, and future possibilities are suggested.  
\end{abstract}

\end{frontmatter}
%
\section{Introduction}

Atomic nuclei are complex many-body systems~\cite{bohr1975,mayer1955} which can be studied by the modern day shell model calculations quite well. However, the dimensions in these modern calculations become quite huge for a full valence space leading to difficulties, particularly for the nuclei near the middle of any given valence space. In spite of being so complicated in nature, nuclear Hamiltonian does support the symmetries in various mass regions resulting in regular features of nuclear spectroscopic properties~\cite{frank2009}. A remarkable example is the occurrence of the seniority scheme in semi-magic (spherical or nearly spherical) nuclei having identical nucleons~\cite{casten1990,jain2021,talmi1993,racah1943,racah1952,shalit1963,heyde1990,isacker2014}. A deviation from seniority behavior may indicate an onset of considerable deformation. This can be of importance in studying $Z=82$, Pb isotopes where shape coexistence is well known~\cite{bengtsson1989,frank2004,heyde2011}. The good seniority states which follow the key features of seniority scheme, can easily be differentiated from other available states. The generalized seniority (GS) scheme is well known in explaining the origin of seniority isomers in the semi-magic nuclei. The GS-suggested multi-j configurations consistently explain the electromagnetic properties such as reduced transition probabilities, half-lives etc. ~\cite{maheshwari2016,maheshwari20161,jain2017,jain20171}. We have extended the GS scheme to obtain the $g-$factor trends by clubbing these GS-suggested multi-j configurations to the Schmidt model and termed it as ‘Generalized Seniority Schmidt Model’(GSSM)~\cite{maheshwari2019}.

The seniority scheme provided by the coupling of valence neutrons may also hold well when few protons are added/removed to the proton closed shell configuration, where these extra proton particles/holes act as spectators for the excited states generated dominantly from the coupling of neutrons. This is indeed the case if two protons are added/removed to the $Z=50$ closed shell resulting in $Z=52$, Te isotopes/ $Z=48$, Cd isotopes, where we have successfully applied the GS scheme to study the various spectroscopic properties such as energies, $B(E2)$ trends, $Q-$moments and $g-$factor trends~\cite{maheshwari20191}. 

The region around $^{208}$Pb is of special interest due to the proximity of two closed shells $Z=82$ and $N=126$ where the structure can be described in terms of few particles/holes coupled to the stable $^{208}$Pb core~\cite{neyens2003}. The experimentally observed $g-$factor trends for the ${13/2}^+$, ${12}^+$ and ${33/2}^+$ isomers in Pb ($Z=82$) isotopes have been explained using GSSM, suggesting them to be generalized seniority $v= 1$, 2 and 3 states, respectively \cite{maheshwari2019}. In this paper, we present a comparative study of the $g-$factor trends for the ${13/2}^+$, ${12}^+$ and ${33/2}^+$ isomers in Hg ($Z=80$, two proton holes) and Po ($Z=84$, two proton particles) isotopes with respect to the Pb ($Z=82$ closed shell) isotopes for the first time using GSSM. The way GSSM has been introduced, it may be considered as a purely phenomenological model which incorporates the effect of multi-j environment as borrowed from generalized seniority. The Hg and Po isotopes have also been extensively studied experimentally due to several interesting structural aspects arising as a consequence of the competition between the proton and neutron pair-breaking and associated multi-quasi particle configurations. Additional arguments in terms of the $Q-$moments and the reduced transition probabilities, $B(E2)$s, have also been given, which also support the GS-suggested multi-j neutron configuration for these ${13/2}^+$, ${12}^+$ and ${33/2}^+$ isomers. Two proton holes/particles in Hg/Po isotopes behave as spectators for these isomers so that the role of neutron configuration is dominant resulting in decay properties very similar to that of Pb isotopes where protons are at the closed shell. In addition, we have studied the first $2^+$ states in Hg, Pb and Po isotopes using GS scheme and explained their inverted parabolic $B(E2)$ trends with respect to increasing neutron number.

We have divided the paper into seven sections. Section 2 presents brief theoretical details and expressions of various spectroscopic properties from the GS scheme for the sake of completeness. Section 3 presents GSSM expressions used in calculating the $g-$factor trends. Section 4 discusses a detailed analysis on the experimental and calculated results for the ${13/2}^+, {12}^+$, and ${33/2}^+$ isomers in Hg, Pb and Po isotopes. Section 5 describes the inverted $B(E2)$ parabolic behavior for the first $2^+$ states in Hg, Pb and Po isotopes. 
Section 6 compiles the future predictions from this phenomenological work. Section 7 presents the conclusions. 

\section{Generalized Seniority Scheme}

Seniority scheme is generally credited to Racah~\cite{racah1943}, who introduced this additional quantum number for differentiating the atomic states having same orbital angular momentum $(L)$, spin angular momentum $(S)$ and total angular momentum $(J)$ quantum numbers. The idea was adopted in nuclear physics in 1952 by Racah~\cite{racah1952} and Flowers~\cite{flowers1952} independently. In simple terms, seniority ($v$) may be defined as the number of unpaired nucleons for a given state. The complete details of seniority scheme in single-j shell may be found in the books of de Shalit and Talmi~\cite{shalit1963}, Casten~\cite{casten1990} and Heyde~\cite{heyde1990}. The description of seniority in quasi-spin algebra, which was originally introduced by Kerman~\cite{kerman1961} and Helmers~\cite{helmers1961}, is on the basis of the pair creation operator $S_j^+$ and pair annihilation operator $S_j^-$ for identical nucleons in single-j shell, where $S_j^+ = \sum_m {(-1)}^{j-m} a_{jm}^+ a_{j,-m}$ and $S_j^-$ is the Hermitian conjugate of $S_j^+$. These pair creation/annihilation operators satisfy the $SU(2)$ Lie algebra; details can be found in the book of Talmi~\cite{talmi1993}. On the basis of Wigner-Eckart theorem in quasi-spin algebra, one can get the seniority reduction formulae for the electromagnetic transitions. For example, when $n$ identical nucleons in a single-j orbital are coupled to give a total spin of $J$, the electromagnetic matrix elements in $j^n$ configuration can be reduced to the electromagnetic matrix elements in $j^v$ configuration. In multi-j shell, the seniority scheme gets extended to the concept of generalized seniority scheme, which was first introduced by Arima and Ichimura~\cite{arima1966} for the multi-j degenerate orbitals. The quasi-spin algebra can now be obtained by defining a generalized pair creation operator $S^+ = \sum_{j} S^+_j$, where the summation over $j$ takes care of the multi-j situation~\cite{talmi1993}. Talmi further incorporated the non-degeneracy of the multi-j orbitals by using $S^+ = \sum_{j} \alpha_j S^+_j$, where $\alpha_j$ are the mixing coefficients for different j-orbitals~\cite{talmi1971,shlomo1972}. 

In this paper, we invoke the GS scheme by defining the quasi-spin operators as $S^+ = \sum_j {(-1)}^{l_j} S^+_j$ ~\cite{arvieu1966}, as also used in our previous papers ~\cite{maheshwari2016,maheshwari20161,jain2017,jain20171,maheshwari2017}. Here $l_j$ denotes the orbital angular momentum of the given-j orbital. The seniority in single-j changes to the generalized seniority $v=\sum_j v_j $ in multi-j. These operators enable us to define a simple pairing Hamiltonian in multi-j shell of various orbitals as $H= 2 S^+ S^- $, which is known to have the energy eigen values $[2s(s+1)-\frac{1}{2} (\Omega-n)(\Omega+2-n)]$ $= \frac{1}{2} [(n-v) (2 \Omega+2-n-v)]$. Here, $s=\sum_j s_j$ is the total quasi-spin of the state having generalized seniority $v=\sum_j v_j $ arising from multi-j $\tilde{j}=j \otimes j' \otimes....$ configuration, with the corresponding pair degeneracy of $\Omega= \sum_j \frac{2j+1}{2}=\frac{2\tilde{j}+1}{2}$. The shared occupancy in multi-j space is akin to the quasi-particle picture. However, the number of nucleons $n=\sum_j n_j$ and the generalized seniority $v=\sum_j v_j$ remain an integer. The pair operators $S^+$ and $S^-$ for multi-j shell, also satisfy quasi-spin $SU(2)$ algebra with generalized seniority as a quantum number. The corresponding selection rules and expressions for the electromagnetic multipole transitions using this GS scheme have been discussed in our earlier works~\cite{maheshwari2016,maheshwari20161,jain2017,jain20171,maheshwari2019,maheshwari20191,maheshwari2017,maheshwarijnp,maheshwari2020,maheshwari20201,agrawal2020} along with firm experimental evidences, wherever available. Kota~\cite{kota2017} has also supported the same using group analogy. For the sake of completeness, below is the list of key features for good generalized seniority states.

\subsection{Key features of good generalized seniority states}

The most prominent signatures of good seniority states show up in the behavior of the excitation energies, reduced electromagnetic transition probabilities such as $B(EL)$ and $B(ML)$ values, $Q-$moments and $g-$factors. We can summarize the features of good generalized seniority states as follows: 

\begin{itemize}

\item[$\bullet$] The excitation energies of good generalized seniority states are expected to have a valence particle number independent behavior, similar to the good seniority states arising from single-j shell~\cite{maheshwari20191}. Consequently, the energy difference remains independent of the valence particle number for a given state in a multi-j configuration. For example, the first excited $2^+$ states in Sn isotopes are observed at nearly constant energy throughout the isotopic chain from $N=52$ to $80$.

\item[$\bullet$] The magnetic dipole moments, i.e. $g-$factors for a given generalized seniority state exhibit a constant trend with respect to particle number variation. In general, the magnetic transition probabilities support a particle number independent behavior for both the even and odd multipole transitions. Therefore, the reduced matrix elements for such transitions between initial $J_i$ and final $J_f$ states can be written as: 
\begin{eqnarray}
\langle \tilde{j}^n v J_f ||\hat{O}(ML)|| \tilde{j}^n v J_i \rangle = \langle \tilde{j}^v v J_f ||\hat{O}(ML)|| \tilde{j}^v v  J_i \rangle
\end{eqnarray}
where $\hat{O}(ML)$ represents the magnetic multipole $(ML)$ operator. This can further result in the particle number independent behavior of magnetic moments for identical nucleons. The matrix elements of magnetic dipole moment $(\hat{\mu})$ in $\tilde{j}^n$ configuration can be reduced to the matrix elements of $\hat{\mu}$ in $\tilde{j}^v$ configuration without $`n $' dependence as follows: 
\begin{eqnarray}
\langle \tilde{j}^n |\hat{\mu} | \tilde{j}^n \rangle = \langle \tilde{j}^v |\hat{\mu}| \tilde{j}^v  \rangle
\end{eqnarray}
We can, therefore, write the magnetic moment of identical nucleons in the multi-j configuration $\tilde{j}^n$ as 
\begin{eqnarray}
\vec{\mu}=g \sum_i^n \vec {\tilde{j_i}} = g\vec{J}
\end{eqnarray}
Here, $\tilde{j}=j \otimes j' \otimes....$ represents the multi-j configuration having the sum of shared occupancies as $n$, the total particle number. Therefore, the $g-$factors $(g)$ of a multi-j configuration also exhibit a particle number independent behavior, similar to the single-j case. Talmi~\cite{talmi1993} has discussed the $g-$factor trend for single-j seniority scheme consisting of $h_{9/2}$ orbital in $N=126$ isotones by fitting the experimental data of magnetic moments. 
In a nutshell, the $g-$factor of all allowed states with generalized seniority $v=2, 3, ...$ arising from a given multi-j configuration must be equal to the $g-$factor of the generalized seniority $v=1$ state arising from the same multi-j configuration. If this is true, the effective interaction will be diagonal in the generalized seniority scheme. 

\item[$\bullet$] The electric transition probabilities exhibit a parabolic behavior for both the odd and even multipole transitions. We recall these developments by the following expressions of electric multipole $L$ (even or odd) operators for the transitions between the initial $J_i$ and final $J_f$ states as:\\
(a)	For generalized seniority preserving ($\Delta v=0$) transitions (when initial $J_i$ and final $J_f$ states are of same generalized seniority, $v$),
\begin{eqnarray}
\langle {\tilde{j}}^n v l J_f ||\sum_i r_i^L Y^{L}(\theta_i,\phi_i)|| {\tilde{j}}^n v l' J_i \rangle = \Bigg[ \frac{\Omega-n}{\Omega-v} \Bigg] \quad \quad \quad \quad && \nonumber \\
\times \langle {\tilde{j}}^v v l J_f ||\sum_i r_i^L Y^{L}(\theta_i,\phi_i)|| {\tilde{j}}^v v l' J_i \rangle && \label{dv0}
\end{eqnarray}
(b)	For generalized seniority changing ($\Delta v=2$) transitions (when initial $J_i$ and final $J_f$ states differ in generalized seniority by 2),
\begin{eqnarray}
\langle {\tilde{j}}^n v l J_f || \sum_i r_i^L Y^{L}(\theta_i,\phi_i)|| {\tilde{j}}^n v\pm 2 l' J_i \rangle  =
  \Bigg[ \sqrt{\frac{(n-v+2)(2\Omega+2-n-v)}{4(\Omega+1-v)}} \Bigg] && \nonumber \\ \langle {\tilde{j}}^v v l J_f ||\sum_i r_i^L Y^{L}(\theta_i,\phi_i)|| {\tilde{j}}^v v\pm 2 l' J_i \rangle \quad &&\label{dv2}
\end{eqnarray} 
where $l$ and $l'$ are the respective parities of final $J_f$ and initial $J_i$ states. Also, $r^L$ and $Y^L$ are, respectively, the radial and spherical harmonic parts of the electric multipole operator. The generalized seniority reduction formulae as given by Eqs. (\ref{dv0}) and (\ref{dv2}), are dominated by total particle number ($n$), total pair degeneracy ($\Omega$) and generalized seniority ($v$). 

\item[$\bullet$] For $L=2$, Eq. (\ref{dv0}) can directly be related to the electric quadrupole moments $ Q= \langle {\tilde{j}}^n J ||\hat{Q}|| {\tilde{j}}^n J \rangle= \langle {\tilde{j}}^n J ||\sum_i r_i^2 Y^{2}|| {\tilde{j}}^n J \rangle $ with the following conclusions: The $Q-$ moment values follow a linear relationship with $n$. The $Q-$values change from negative to positive on filling up the given multi-j shell, with a zero value in the middle of the shell due to $\frac{\Omega-n}{\Omega-v}$ term. This is in direct contrast to the $Q-$moment generated by collective deformation which is expected to be the largest in the middle of the shell. This is how we can differentiate between the single-particle and collective excitations. 

\item[$\bullet$] The dependence of the $\sqrt{B(E2)}$ with particle number $n$ in Eq. (\ref{dv2}) for the generalized seniority changing transitions is different than the case of $Q-$moments. The $\sqrt{B(E2)}$ values for $\Delta v=2$ transitions exhibit a flat trend throughout the multi-j shell, decreasing to zero at both the shell boundaries. A nearly spherical structure is supported at both the ends for the given multi-j shell.

\item[$\bullet$] The first excited $2^+$ states with generalized seniority $v=2$ usually decay to the ground $0^+$ states, a fully pair-correlated state with generalized seniority $v=0$. Such $E2$ transitions are the generalized seniority changing transitions ($\Delta v=2$), where the corresponding $B(E2)$ values can be obtained as follows: 
\begin{eqnarray}
B(E2)=\frac{1}{(2J_i+1)}| \langle {\tilde{j}}^n v l J_f || \sum_i r_i^2 Y^{2}(\theta_i,\phi_i) || {\tilde{j}}^n v\pm2 l' J_i \rangle |^2
\end{eqnarray}

\end{itemize}

\section{The Generalized Seniority Schmidt model and the $g-$factors}

Nuclear moment measurements are difficult in comparison to the measurement of other nuclear properties; however, the moments can reveal information which is usually not accessible by other properties. Such valuable measurements further allow confirmation of ideas based on direct/indirect experimental evidences. They also provide crucial inputs in nuclear models for testing model parameters, appropriateness of the used parameterization and model space. The nuclear magnetic moments provide a good test for the purity of a particular configuration since they are sensitive to the orbits occupied by the valence nucleons, among which, they are most sensitive to the orbits with unpaired nucleons. However, the magnetic moments are very little sensitive to the orbits with paired nucleons as long as these nucleons are fully coupled to zero spin. Keeping this in mind, we have recently developed a new model \cite{maheshwari2019} to calculate the $g-$factors (i.e. magnetic moments) by merging the idea of generalized seniority (based on the orbits with unpaired nucleons) with the well-known Schmidt model \cite{schmidt1937} and calling it as `Generalized Seniority Schmidt Model' (GSSM). The GSSM expressions \cite{maheshwari2019} are obtained by extending the Schmidt model of single-j to the effective multi-j $\tilde{j}=j \otimes j' \otimes....$, and can be written as :
\begin{eqnarray}
g  =& \frac{1}{\tilde{j}} \Bigg[ {\frac{1}{2} g_s}+ (\tilde{j}- \frac{1}{2}) g_l \Bigg]; \tilde{j}=\tilde{l}+\frac{1}{2}\nonumber\\ 
g  =& \frac{1}{\tilde{j}+1} \Bigg[ -\frac{1}{2} g_s + (\tilde{j}+ \frac{3}{2}) g_l \Bigg]; \tilde{j}=\tilde{l}-\frac{1}{2} 
\end{eqnarray}

where $g_s$ and $g_l$ are taken to be 5.59 $n.m$. and 1 $n.m.$ for protons, while $-3.83$ $n.m.$ and 0 $n.m.$ for neutrons, respectively. This is a purely phenomenological adoption of generalized seniority, which seems to work reasonably well. The GSSM calculated results come closer to the experimental data than the pure Schmidt model (single-j) for various semi-magic seniority isomers in Sn and Pb isotopes, and $N=82$ isotones~\cite{maheshwari2019}. Interestingly, the GSSM works quite well in explaining the $g-$factor trends of ${11/2}^-$ isomers in both Cd and Te isotopes (away from semi-magicity) in a way similar to the Sn isotopes ~\cite{maheshwari20191}. This may be due to the fact that the multi-j environment in GSSM takes care of the spin quenching of $g_s$ by the amount of $(j / \tilde j)$ in comparison to the Schmidt model. However, the contribution of $g_l$ in multi-j GSSM is almost similar to the case of single-j Schmidt model. This multi-j environment along with the proper configuration mixing of valence nucleons with quasi-particle picture on the magnetic moment operator results in an overall treatment of spin quenching as in any first order perturbation theory~\cite{zamick1971,nomura1972}. The corresponding $g-$factor trends hence come closer to the experimental data qualitatively. Further quantitative matching will indeed require the treatment of higher-order microscopic effects as in other microscopic methods involving core polarisation or meson exchange currents etc.~\cite{towner1987}. Interestingly, GSSM results do not need any kind of tuning to estimate the amount of spin quenching to explain the experimental data. The spin quenching is governed by the multi-j configuration $(\tilde j)$ in GSSM as suggested by generalized seniority, which consistently explains other nuclear properties. In this paper, we have applied GSSM to study the $g-$factor trends in Hg and Po isotopes, having two holes and two particles, respectively, to the $Z=82$ proton closed-shell of Pb isotopes. 

\section{Generalized seniority isomers in Hg, Pb and Po isotopes }

We now present the generalized seniority results for the ${13/2}^+$, ${12}^+$ and ${33/2}^+$ isomers in Hg, Pb and Po isotopes. The pair degeneracy $\Omega= 13$ corresponds to the multi-j \{$i_{13/2} \otimes f_{7/2} \otimes p_{3/2} $\} neutron configuration in the following discussion. This configuration has been chosen by freezing the h$_{9/2}$ orbital of $N=82-126$ valence space so that the active valence space varies from $N=92-118$. This multi-j \{$i_{13/2} \otimes f_{7/2} \otimes p_{3/2} $\} configuration has been considered to be originating from $ \tilde{j}=\tilde{l}+\frac{1}{2}$, since $i_{13/2}$ plays the dominating role. Several of the spin-parity assignments for these ${13/2}^+$, ${12}^+$ and ${33/2}^+$ states in Hg, Pb and Po isotopes are not yet confirmed \cite{ensdf} but our calculated results support these systematic assignments in all the three isotopic chains. The GS calculations for $Q-$moments and $B(E2)$ trends are performed from Eqs. (4) and (5), and using the value obtained by fitting one of the experimental data. This single-point fitting takes care of the involved structural effects such as single-particle matrix elements, radial integrals etc. Though, GSSM formulas in Eq. (7) do not require any such fitting. 
\begin{figure}[!htb]
\includegraphics[width=13cm,height=9cm]{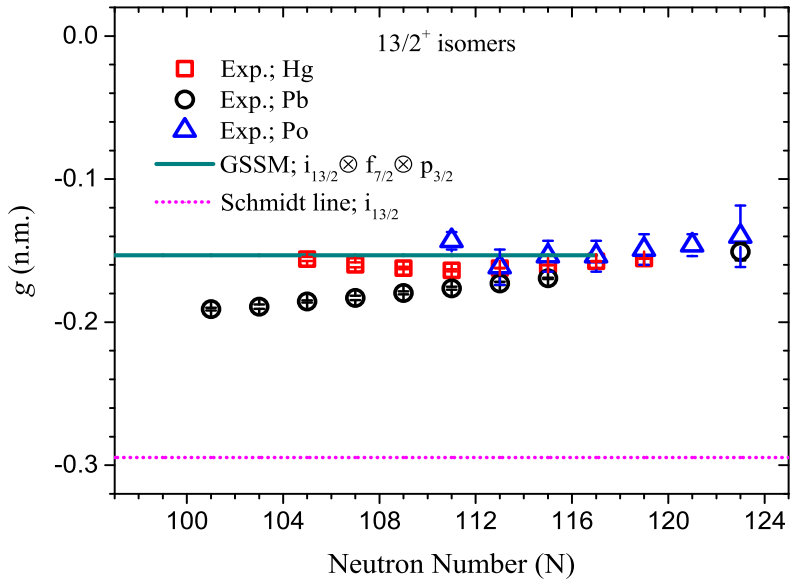}
\caption{\label{fig:g13}(Color online) Experimental~\cite{stone2019,stone2014} and GSSM calculated $g-$factor trends for the ${13/2}^+$ isomers in Hg, Pb and Po isotopes. The calculations have been done using $\Omega=13$ corresponding to $v=1$, \{$\tilde{j}=i_{13/2} \otimes f_{7/2} \otimes p_{3/2} $\} multi-j configuration. Schmidt line for pure neutron $i_{13/2}$ orbital is also shown for comparison.}
\end{figure}

\begin{figure}[!htb]
\includegraphics[width=12cm,height=9cm]{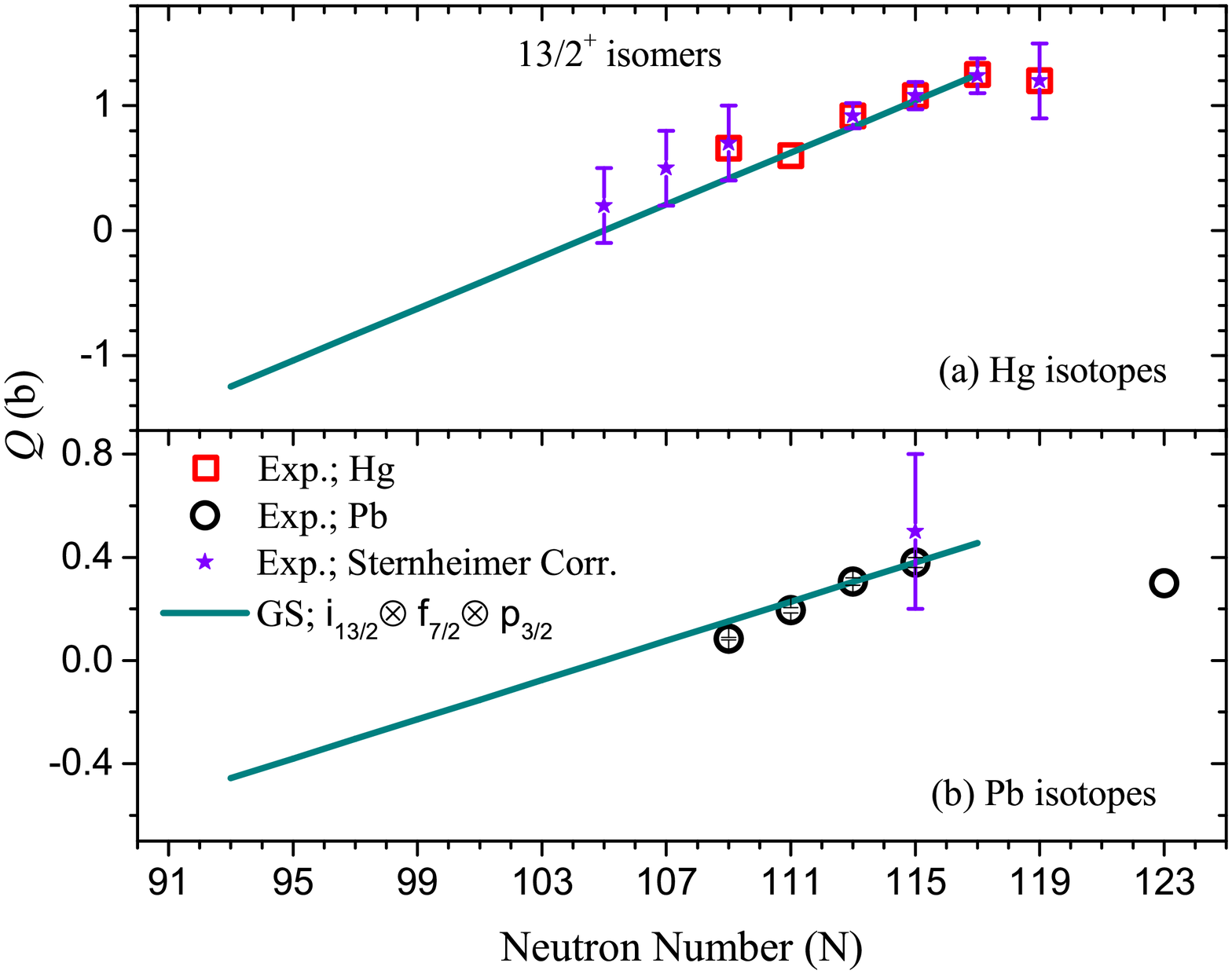}
\caption{\label{fig:q13}(Color online) Experimental~\cite{stone2014} and calculated Q-moment trends for the ${13/2}^+$ isomers in Hg and Pb isotopes. The calculations have been done using $\Omega=13$ corresponding to $v=1$, \{$\tilde{j}=i_{13/2} \otimes f_{7/2} \otimes p_{3/2} $\} multi-j configuration. In case of multiple measurements, the weighted averaged value has been adopted. Besides this, Sternheimer shielding corrected values are also shown, wherever available. Refer text for more details. } 
\end{figure}

\subsubsection{${13/2}^+$ isomers}

\noindent \textit{$g-$factor} \\
The ${13/2}^+$ isomers are very long-lived having half-lives in minutes and occur at about 100 keV above the respective ground state in odd-A Hg, Pb and Po isotopes. We present in Fig.~\ref{fig:g13}, the measured $g-$factor trends of these ${13/2}^+$ isomers in all the three Hg ($Z=80$), Pb ($Z=82$) and Po ($Z=84$) isotopic chains. The measured values have been taken from the latest update of Stone's table in 2019 \cite{stone2019}. For those nuclei in which the measured $g-$factors are not available in this latest update due to half-life limit or other constraints, we have adopted the data from Stone's earlier compilation in 2014 \cite{stone2014} to complete the systematics; follow Table~\ref{tab:expdata} for the measured values shown in Fig.~\ref{fig:g13}. We can clearly see a particle number independent behavior of $g-$factors for the ${13/2}^+$ states, as expected from the seniority and generalized seniority scheme, changing not more than $10\%$ around an average value. 

We calculate the $g-$factor trend for these $v=1$, ${13/2}^+$ states by the GSSM using $\Omega=13$ multi-j configuration. The GSSM calculated results explain the experimental data quite well supporting the quasi-particle nature of odd-neutron involved. The ${13/2}^+$ isomers are expected to be pure seniority states with $i_{13/2}$ unique-parity orbital configuration of the $82-126$ neutron valence space. However, it seems to have a multi-j wave function since the Schmidt $g-$factors for $i_{13/2}$ neutrons come out to be $-0.2946$ $n.m.$, which lies quite far from the experimental values as shown in Fig.~\ref{fig:g13}. This strongly hints towards a picture having shared occupancy of generalized seniority $v=1$ from \{$\tilde{j}=i_{13/2} \otimes f_{7/2} \otimes p_{3/2} $\} configuration for the ${13/2}^+$ isomers in all the three Hg, Pb and Po isotopic chains. 

Wouters $et$ $al.$ \cite{wouters1991} explained the systematic increase of measured $g-$factor values as a function of neutron number in the Pb isotopes, in terms of an increasing first-order core polarization for pure i$_{13/2}$ orbital. Also, they attributed the large deviation of $g-$factor with respect to the Schmidt value ($-0.2946$ n.m.) in Hg and Po isotopes, by using the increase of second-order core polarization in the non-magic nuclei \cite{neyens2003,wouters1991}. In our calculations, we are simply able to explain this deviation of $g-$factor from Schmidt value by using the generalized seniority multi-j configuration in spite of pure-j i$_{13/2}$ orbital. The GSSM line falls very close to the experimental data in comparison to the pure Schmidt line. This deviation may qualitatively be related to the multi-j occupancy rather than pure i$_{13/2}$ occupancy; however, the i$_{13/2}$ orbital is the dominating contributor. One further needs the inclusion of fine structural effects as used in the other microscopic methods for a quantitative explanation.
\\
\\
\textit{$Q-$moment} \\
We now discuss the quadrupole $(Q)-$moment trends of these ${13/2}^+$ isomers in Hg, Pb and Po isotopes. The spectroscopic quadrupole moment of a nuclear state with spin measures the deviation of the nuclear charge distribution from sphericity. While the nuclear magnetic moment is very sensitive to the single-particle orbits occupied by the unpaired nucleons, the $Q-$moment is a unique tool to study the deformation and collective behavior of nuclei. Fig. \ref{fig:q13} presents the comparison of experimental and calculated $Q-$moment trends for the ${13/2}^+$ isomers in Hg and Pb isotopes. All the experimental data have been taken from Stone's 2014 compilation \cite{stone2014}. In case of multiple measurements available for a given isotope such as $^{191}$Hg and $^{195}$Pb, the weighted average value has been adopted in Fig. \ref{fig:q13}. The sign of the measured $Q-$value for $^{205}$Pb is not known and assumed to be positive. No experimental data are available for the $Q-$moments of these isomers in Po isotopes. For most of the measured $Q-$moments in Hg isotopes and $^{197}$Pb isotope, Sternheimer shielding correction has been made by the original authors as cited in~\cite{stone2014}, which have also been shown in Fig. \ref{fig:q13} for comparison and to complete the trend; as mentioned in Table~\ref{tab:expdata}. 

The middle of active neutron valence space $(92-118)$ lies at $N=105$. As expected from the GS scheme, the calculated $Q-$moment increases linearly with increasing neutron number. It starts with a negative $Q-$value at $N=93$, reaches to zero at the middle $N=105$, and then becomes positive for $N>105$. The calculations have been done using $\Omega=13$ and generalized seniority $v=1$ configuration for these isomers. The experimental data are not available below $N=105$ for any of the isotopic chains since these nuclei are quite far from stability and are difficult to measure. The GS calculated $Q-$moment trend reproduces the measured values quite well for Pb and Hg isotopes, where experimental data are available. In $N=105,107,109$, Hg isotopes, the GS calculated line remains within the experimental error bars.  After $N=118$, the measured $Q-$moment starts to fall clearly supporting the change in multi-j configuration which is beyond the scope of $\Omega=13$. 

Neyens~\cite{neyens2003} explicitly discussed the spherical shell model picture for these ${13/2}^+$ isomers in Pb isotopes from $N = 106$ to 120, where the number of nucleons in the $i_{13/2}$ orbital is $n = 1$ at $N = 107$ to $n = 13$ at $N = 119$. However, one can not explain the occurrence of ${13/2}^+$ isomers below $N=107$ in Hg and Pb isotopes using the single-j picture. Our GS results use the multi-j neutron valence space from $N = 92$ to $118$ where number of nucleons in multi-j \{$\tilde{j}=i_{13/2} \otimes f_{7/2} \otimes p_{3/2} $\} orbitals is $n=1$ at $N=93$ to $n=25$ at $N=117$. This also explains the anomaly of all positive $Q-$value for these ${13/2}^+$ isomers in Hg and Pb isotopes ~\cite{neyens2003} since the middle occur at $N=105$ and no measured $Q-$moment value exists for very neutron-deficient isotopes with $N < 105$. More experimental efforts are hence needed. The multi-j configuration to explain $Q-$moments of the ${13/2}^+$ isomers is consistent with the configuration used in GSSM, for $g-$factor trend line of these isomers. Hence, the ${13/2}^+$ isomers can be understood as the generalized seniority $v=1$ states from \{$\tilde{j}=i_{13/2} \otimes f_{7/2} \otimes p_{3/2} $\} configuration. As soon as we cross the $N=118$, the $Q-$moment values change the slope and start to decrease with an increase in neutron number, which clearly suggests the change in orbitals, or the saturation of $i_{13/2}$ occupancy in total wave functions.  

\subsubsection{${12}^+$ isomers}
\begin{figure}[!htb]
\includegraphics[width=13cm,height=11cm]{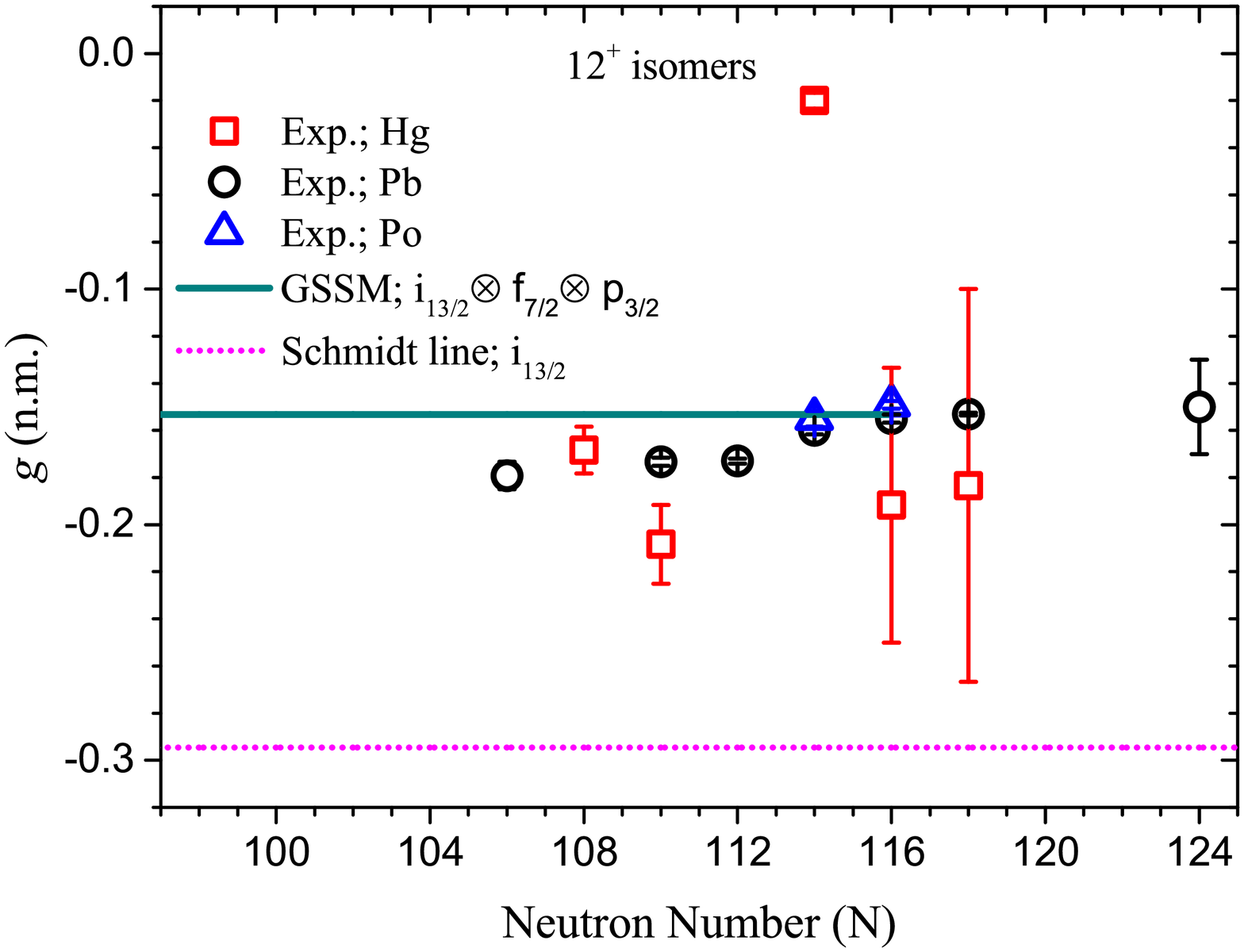}
\caption{\label{fig:g12}(Color online) Experimental~\cite{stone2014} and GSSM calculated $g-$factor trends for the ${12}^+$ isomers in Hg, Pb and Po isotopes. The calculations are done using $\Omega=13$ corresponding to $v=2$, \{$\tilde{j}=i_{13/2} \otimes f_{7/2} \otimes p_{3/2} $\} multi-j configuration. Schmidt line for pure neutron $i_{13/2}$ orbital is also shown for comparison.} 
\end{figure}

\noindent \textit{$g-$factor} \\
The ${12}^+$ isomers are short-lived having a half-life of around 100 $ns$ and at excitation energy of about 2 MeV which is needed to break up the neutron pair. Fig.~\ref{fig:g12} presents a comparison of experimental and GSSM calculated $g-$factor trends of ${12}^+$ isomers in Hg, Pb and Po isotopes. A nearly constant $g-$factor trend can be seen for all the three isotopic chains. For $^{196}$Hg and $^{192,198,200}$Pb isotopes, the weighted average of the measured $g-$factor values has been taken~\cite{stone2014}; follow Table~\ref{tab:expdata12} for the measured values. The calculated results support the generalized seniority $v=2$, neutron configuration for these states using $\Omega=13$, except for $^{190,194}$Hg at $N=110,114$. The measured $g-$factor value for $^{194}$Hg at $N=114$ is taken out to be averaged since several states were unresolved for this nucleus; refer to Stone's compilation for more details \cite{stone2014}. These measurements have been done in 1980s and need further experimental efforts for precise values. In Po isotopes, only two measured values for $^{198,200}$Po at $N=114,116$ are available and found to be in good match with the GSSM line. Future experiments may complete the systematics for a better clarity. 
\\
\begin{figure}[!htb]
\includegraphics[width=13cm,height=11cm]{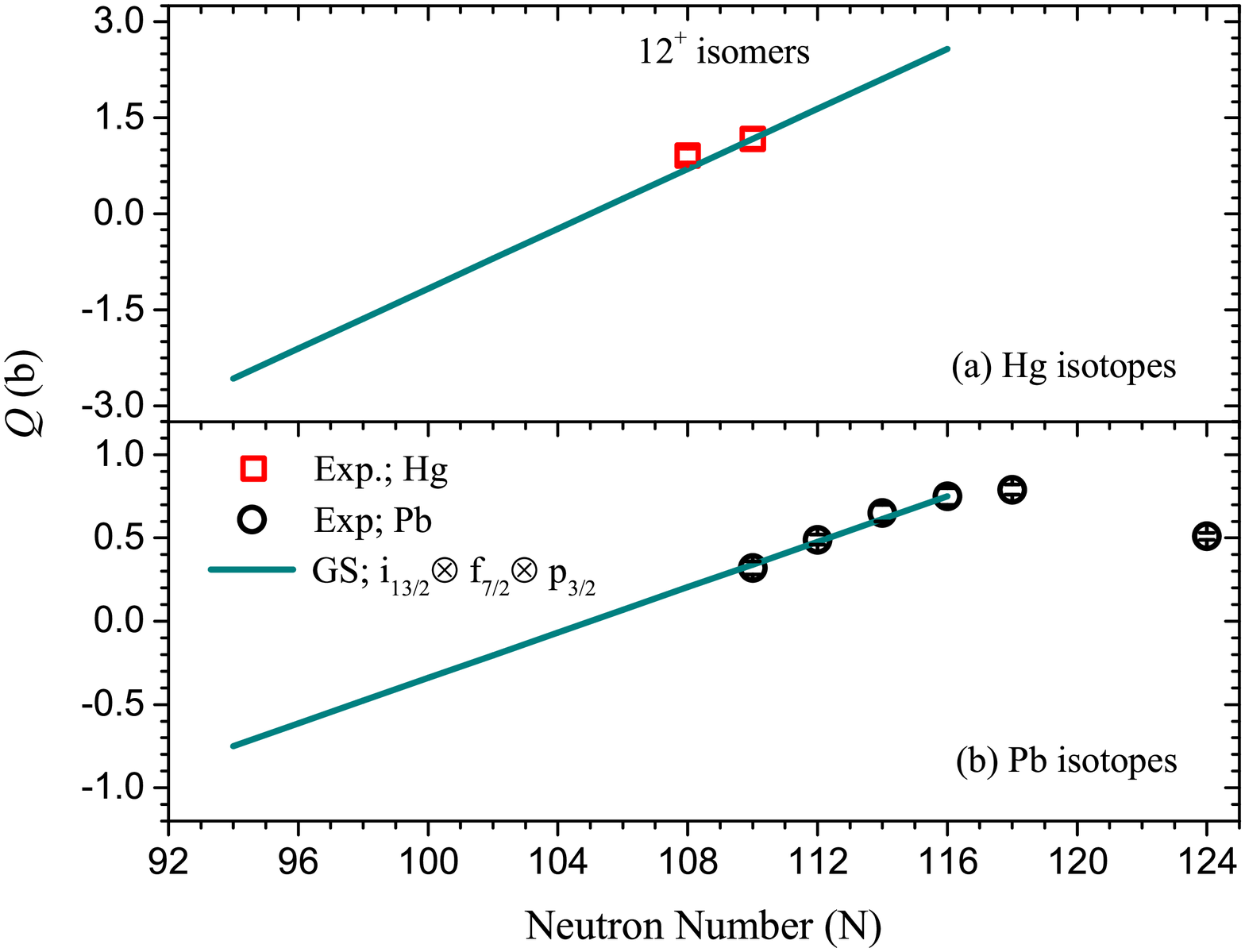}
\caption{\label{fig:q12}(Color online) Experimental~\cite{stone2014} and calculated $Q-$moment trends for the ${12}^+$ isomers in Hg, Pb and Po isotopes. The calculations are done using $\Omega=13$ corresponding to $v=2$, \{$\tilde{j}=i_{13/2} \otimes f_{7/2} \otimes p_{3/2} $\} multi-j configuration.} 
\end{figure}
\\
\textit{$Q-$moment} \\
Fig.~\ref{fig:q12}(a) and \ref{fig:q12}(b) present a comparison of experimental ~\cite{stone2014} and GS calculated $Q-$moment trends of ${12}^+$ isomers in Hg and Pb isotopes, respectively. Very limited data are available for comparison and all being without sign. In $^{206}$Pb, the measured $Q-$value was estimated from the $B(E2; {12}^+ \rightarrow {10}^+)$ value of ${12}^+$ isomer by Mahnke $et$ $al.$ ~\cite{mahnke1979}. Table~\ref{tab:expdata12} lists the measured values adopted in Fig.~\ref{fig:q12}. The calculated results support GS $v=2$ neutron configuration for these isomers using $\Omega=13$ (\{$\tilde{j}=i_{13/2} \otimes f_{7/2} \otimes p_{3/2} $\}) and are in line with the measured values (assumed to be positive), wherever available. As expected from GS scheme, the calculated $Q-$moment increases linearly by increasing neutron number. It starts with a negative $Q-$value at $v=2, N=94$, reaches to zero at the middle, and then becomes positive for $N>105$. The measured $Q-$moment value at $^{200}$Pb seems to be originating from the saturated wave functions involving $i_{13/2}$ beyond the scope of given $\Omega=13$ configuration. No $Q-$moment measurements are available for Po isotopes.
\\
\begin{figure}[!htb]
\includegraphics[width=13cm,height=11cm]{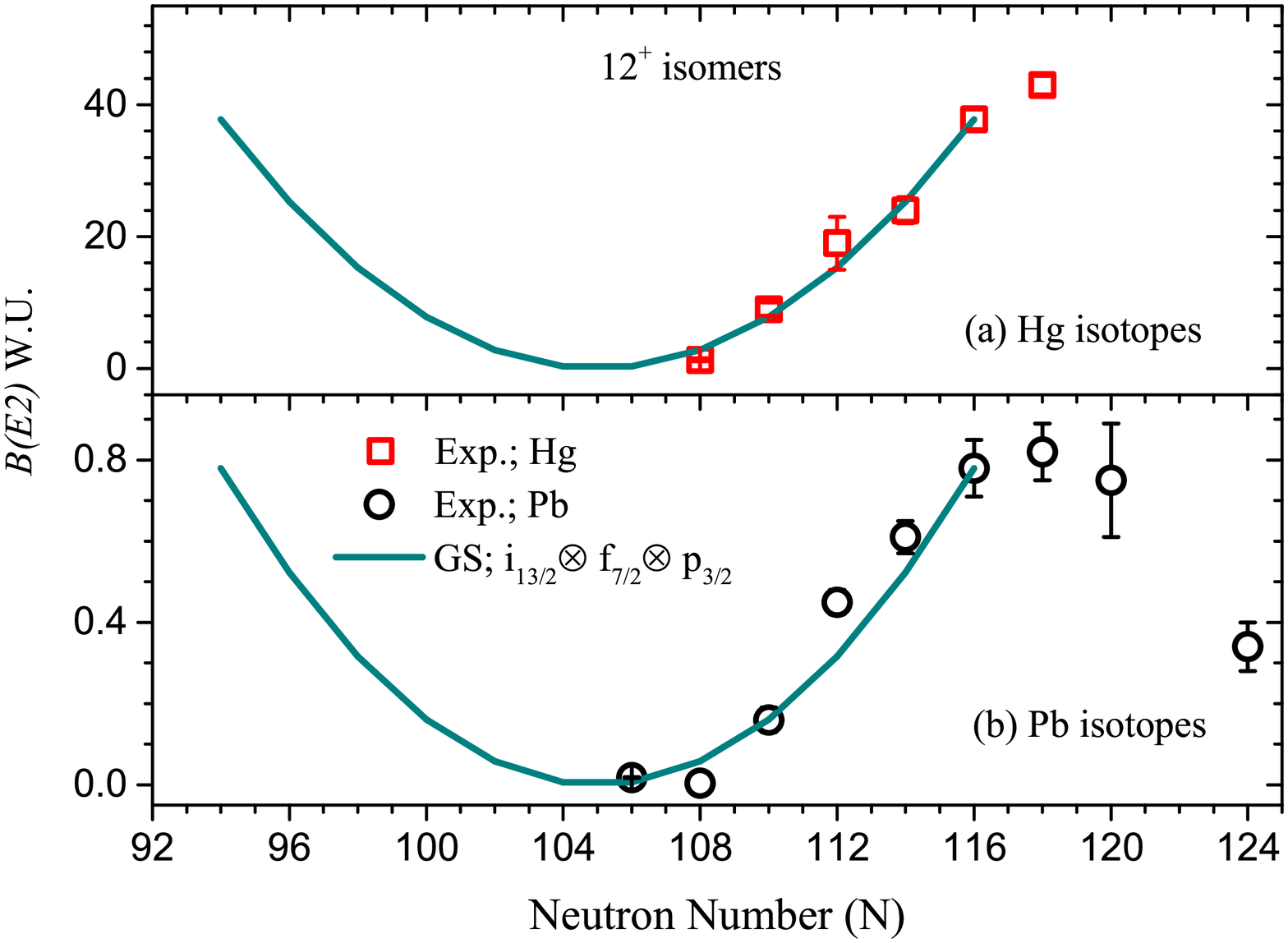}
\caption{\label{fig:be212}(Color online) Experimental \cite{ensdf} and calculated $B(E2)$ trends for the ${12}^+$ isomers in Hg and Pb isotopes. For $^{190}$Pb ($N=108$), the measured $B(E2)$ value has been estimated considering it to be decaying by $\sim$ 120 keV gamma; refer text for more details.} 
\end{figure}
\\
\textit{$B(E2)$ trend} \\
Figs. \ref{fig:be212}(a) and \ref{fig:be212}(b) present the experimental and GS calculated $B(E2)$ trends in Weisskopf Units (W.U.) for these ${12}^+$ isomers in Hg and Pb isotopes, respectively. The GS calculated trends are able to explain the measured $B(E2)$ values using $v=2, \Omega=13$ configuration for these isomers up to $N=116$. As soon as one crosses the $N=118$, another $B(E2)$ parabola seems to appear due to a change of multi-j configuration having a saturated occupancy of $i_{13/2}$ orbital and the remaining $f_{5/2}$ and $p_{1/2}$ orbitals from $N=82-126$ valence space. In $N=108$, $^{190}$Pb, the experimental $B(E2)$ value in Fig. \ref{fig:be212}(b) has been obtained using the reported half-life value, assuming the state to be decaying by $E2$ transition. Since the $\gamma-$decay is unobserved in this case, $E_{\gamma}=120$ keV ~\cite{dracoulis1998} has been taken to calculate the $B(E2)$ value. Internal conversion coefficient has been calculated using BRICC calculator~\cite{bricc}. Hence, the value is shown without error. Both spin and parity of this ${12}^+$ state in $^{190}$Pb are not yet confirmed experimentally~\cite{ensdf}. In $^{204}$Hg $(N=124)$, only an upper limit on half-life of the ${12}^+$ isomer as $< 5 $ ns is known due to the life-time sensitivity of the experiment~\cite{wrzesinski2015}. Table~\ref{tab:expdata12} refers to the list of the measured $B(E2)$ values shown in Fig.~\ref{fig:be212}. No experimental data for these ${12}^+$ isomers are available in Po isotopes. The ${12}^+$ isomers in Hg isotopes can hence be understood as $v=2$ isomers arising from \{$\tilde{j}=i_{13/2} \otimes f_{7/2} \otimes p_{3/2} $\} configuration similar to those in Pb isotopes.
  
\subsubsection{${33/2}^+$ isomers}

\begin{figure}[!htb]
\includegraphics[width=14cm,height=11cm]{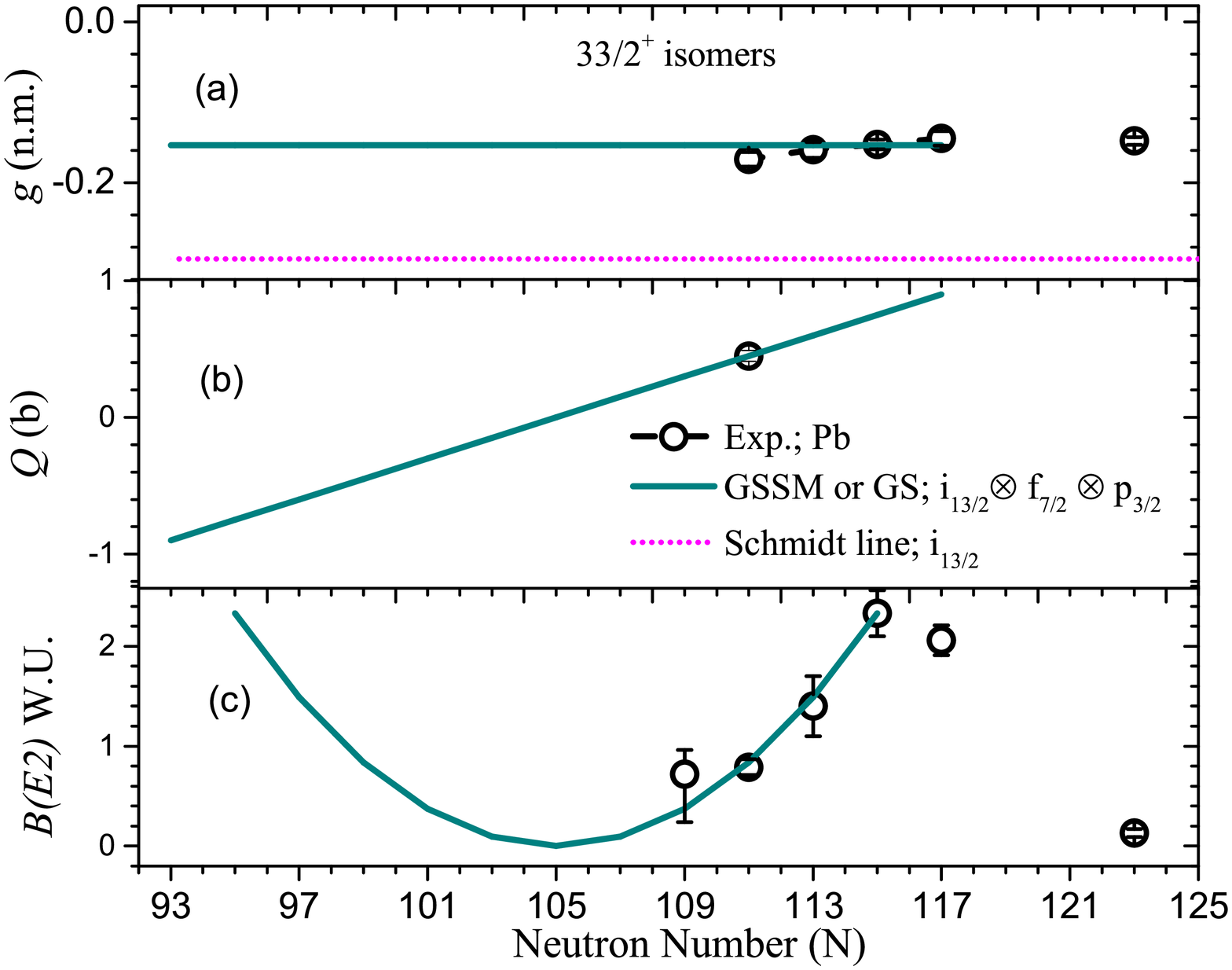}
\caption{\label{fig:merged33}(Color online) Experimental~\cite{stone2014} and calculated trends for the $g-$factors, $Q-$moments and $B(E2)$ values of the ${33/2}^+$ isomers in Pb isotopes. The calculations have been done using $\Omega=13$ corresponding to $v=3$, \{$\tilde{j}=i_{13/2} \otimes f_{7/2} \otimes p_{3/2} $\} multi-j configuration. The measured values for moments (i.e. $g-$factors, panel (a) and $Q-$moments, panel (b)) have been taken from the Stone's compilation in 2014 \cite{stone2014}. The measured $B(E2)$ values, panel (c), have been adopted from ENSDF \cite{ensdf}.} 
\end{figure}

\noindent \textit{$g-$factor} \\
Fig. \ref{fig:merged33}(a) represents the experimental and calculated $g-$factor trends for the ${33/2}^+$ isomers in Pb isotopes. The calculations are done using $\Omega=13$ and generalized seniority $v=3$. The measured values have been taken from Stone's compilation in 2014 \cite{stone2014}. In $^{195}$Pb, the experimental weighted averaged $g-$factor value has been adopted due to two available measurements, as given in \cite{stone2014}. All the measured values shown in the figure are listed in Table~\ref{tab:expdata33}. The calculated GSSM line explains the measured values quite well exhibiting a particle number independent behavior. Since $v=1,$ ${13/2}^+$ isomers are of similar nature in all the three Hg, Pb and Po isotopic chains, it is highly expected to have similar $g-$factor trends for the ${33/2}^+$ isomers in Hg and Po isotopes, where no $g-$factor data are available till date. New moment measurements may confirm the situation. \\
\\
\textit{$Q-$moment} \\
We have also calculated the $Q-$moment trend of these ${33/2}^+$ isomers in Pb isotopes using GS scheme, where only a single measurement is available till date \cite{stone2014}, as shown in Fig \ref{fig:merged33}(b) and listed in Table~\ref{tab:expdata33}. The calculations are done using $\Omega=13$ and $v=3$ configuration. No $Q-$moment measurements are available for these isomers in Hg and Po isotopes. It would be interesting to have more experimental measurements for Pb as well as for Hg and Po isotopes to get a more clear picture for this $v=3$ isomer.
\\
\\
\textit{$B(E2)$ trend} \\
Fig. \ref{fig:merged33}(c) presents the measured and GS calculated $B(E2)$ trends in W.U. for these ${33/2}^+$ isomers in Pb isotopes. The measured $B(E2)$ values have been adopted from the respective ENSDF (Evaluated Nuclear Structure Data File) data base \cite{ensdf} and are listed in Table~\ref{tab:expdata33}. The GS trend explains the measured values quite well, wherever available, up to $N=115$ that is the three-hole configuration in the active valence space as defined by $\Omega=13$. The measured $B(E2)$ values start to decrease on crossing $N=115$ which is very similar to the behavior of ${12}^+$ isomers on crossing $N=116$ in even-A Pb isotopes. This may again be understood as a change in multi-j configuration leading to the saturated $i_{13/2}$ occupancy for these higher-mass Pb isotopes towards the next neutron closed shell. One can also predict the occurrence of ${33/2}^+$ isomers in lighter-mass Pb isotopes following the calculated GS trend shown in Fig. \ref{fig:merged33}(c). 

Similar situation may be expected for these isomers in Hg and Po isotopes. The ${33/2}^+$ states are observed in Hg isotopes with $N=99-117$; however, the life-times are not available for most of them except $^{191,193}$Hg at $N=111,113$. In $^{195,197}$Hg at $N=115,117$, the lower limits on half-lives are also known. Though the current situation is not in-line with seniority predictions, one needs more precise measurements to conclude on the origin of these ${33/2}^+$ isomers in Hg isotopes. Very limited experimental data are available for these $v=3$, ${33/2}^+$ states in Po isotopes as well.   

\begin{table*}[!htb]
\caption{\label{tab:expdata} The experimental data for the $g-$factor and $Q-$moment trends of the ${13/2}^+$ isomers in Hg, Pb and Po isotopes. The $g-$factor data have been adopted from Stone's latest compilation in 2019~\cite{stone2019} while the $Q-$moment data have been taken from Stone's compilation in 2014~\cite{stone2014}, unless otherwise stated. The uncertainties are shown in the parentheses. In case of multiple measurements, the weighted average value has been adopted. Sternheimer corrected (St. corr.)  $Q-$ moment values are also shown for comparison.}
\begin{center}
\begin{tabular}{c c c c c}
\hline
N & Nucleus & $g-$factor (n.m.)~\cite{stone2019} & \multicolumn{2}{c}{$Q-$moment (b)~\cite{stone2014}}  \\
& & & & St. corr. \\
\hline 
105	& $^{185}$Hg & -0.1559(14) & - & 0.2(3) \\ 
107	& $^{187}$Hg & -0.1600(17) & - & 0.5(3) \\
109	& $^{189}$Hg & -0.1622(9) & 0.66(19) & 0.7(3)   \\
111	& $^{191}$Hg & -0.1637(8) & 0.6(2) &-   \\
113	& $^{193}$Hg & -0.1622(2) & 0.92(2) & 0.92(10)   \\
115	& $^{195}$Hg & -0.1601(2) & 1.08(2) & 1.08(11)    \\
117	& $^{197}$Hg & -0.1575(2) & 1.25(3) & 1.24(14)    \\
119	& $^{199}$Hg & -0.1555(2) & 1.2(3) & 1.2(3)  \\
\hline
101	& $^{183}$Pb & -0.1909(9) & - & - \\
103	& $^{185}$Pb & -0.1892(15) & - & - \\
105 & $^{187}$Pb & 	-0.1855(8) & - & - \\
107	& $^{189}$Pb & 	-0.1831(15) & - & - \\
109	& $^{191}$Pb & 	-0.1795(11) & 0.085(5) & - \\
111	& $^{193}$Pb & 	-0.1763(11) &  0.195(10) & - \\
113	& $^{195}$Pb &	-0.1729(11) &  0.306(15) & - \\
115	& $^{197}$Pb & 	-0.1694(4) &  0.38(2) & 0.5(3) \\
123	& $^{205}$Pb & -0.151(6) \cite{stone2014} & 0.30(5) & - \\
\hline 
111	& $^{195}$Po &	-0.143(6)  & - & - \\
113	& $^{197}$Po &	-0.162(12)  & - & - \\
115	& $^{199}$Po &	-0.154(11)  & - & - \\
117	& $^{201}$Po &	-0.154(11)  & - & - \\
119	& $^{203}$Po &	-0.149(11)  & - & - \\
121	& $^{205}$Po &	-0.146(7) \cite{stone2014} & - & - \\
123	& $^{207}$Po &	-0.140(21) \cite{stone2014} & - & - \\
\hline
\end{tabular}
\end{center}
\end{table*}

\begin{table*}[!htb]
\caption{\label{tab:expdata12} The experimental data for the $g-$factor, $Q-$moment and $B(E2)$ trends of the ${12}^+$ isomers in Hg, Pb and Po isotopes. The $g-$factor data have been adopted from Stone's latest compilation in 2019~\cite{stone2019} while the $Q-$moment data have been taken from Stone's compilation in 2014~\cite{stone2014}, unless otherwise stated. The measured $B(E2)$ values are taken from ENSDF \cite{ensdf}. The uncertainties are shown in the parentheses. In case of multiple measurements, the weighted average value has been adopted. }
\begin{center}
\begin{tabular}{c c c c c}
\hline
N & Nucleus & $g-$factor (n.m.)~\cite{stone2019} & $Q-$moment (b)~\cite{stone2014} & $B(E2)$ W.U.~\cite{ensdf} \\
\hline 
108 & $^{188}$Hg &	-0.168(10) & 0.91(11) & 1.30(23) \\ 
110	& $^{190}$Hg & -0.208(17) & 1.17(14) & 9(1)  \\ 
112	& $^{192}$Hg & - & - & 19(4)  \\ 
114	& $^{194}$Hg & -0.020(3)~\cite{stone2014} & - & 24(2)  \\ 
116	& $^{196}$Hg & -0.19(6) & - & 37.8(15)  \\ 
118 & $^{198}$Hg &	-0.18(8) & - &  43.0(14)  \\ 
\hline
106 & $^{188}$Pb & -0.179(6) & - & 0.0177(15) \\ 
108	& $^{190}$Pb & -0.168(10) & - & 0.00358* \\ 	
110	& $^{192}$Pb & -0.173(2) & 0.32(4) & 0.16(3) \\ 
112	& $^{194}$Pb & -0.173(1) & 0.49(3) & 0.45(3)  \\ 
114	& $^{196}$Pb & -0.160(2) & 0.65(5) & 0.61(4)  \\ 
116	& $^{198}$Pb & -0.155(2) & 0.75(5) & 0.78(7)  \\ 
118	& $^{200}$Pb & -0.1530(6) & 0.79(3) & 0.82(7)  \\ 
120	& $^{202}$Pb & - & - & 0.75(14)  \\ 			
122	& $^{204}$Pb & - & - & - \\ 			
124	& $^{206}$Pb & -0.15(2) & 0.51(2)~\cite{mahnke1979} & 0.34(6)  \\ 
\hline 
114 & $^{198}$Po & -0.155(3) & - & - \\ 
116 & $^{200}$Po & -0.1491(17) & - & -  \\
\hline
\multicolumn{5}{l}{* Calculated using $E_{\gamma}=120$ keV ~\cite{dracoulis1998}}
\end{tabular}
\end{center}
\end{table*}

\begin{table*}[!htb]
\caption{\label{tab:expdata33} Same as Table \ref{tab:expdata12}, but for the ${33/2}^+$ isomers in Pb isotopes. }
\begin{center}
\begin{tabular}{c c c c c}
\hline
N & Nucleus & $g-$factor (n.m.)~\cite{stone2019} & $Q-$moment (b)~\cite{stone2014} & $B(E2)$ W.U.~\cite{ensdf} \\
\hline
109	& $^{191}$Pb & 	- & -  & 0.72 ($^{+24}_{-48}$) \\
111	& $^{193}$Pb & 	-0.171(9) & 0.45(4) & 0.79(8) \\
113	& $^{195}$Pb & -0.160(10) &  - & 1.4(3) \\
115	& $^{197}$Pb & 	-0.152(6) & -  & 2.33(23)  \\
117	& $^{199}$Pb & 	-0.145(9) &  - & 2.06(15) \\
123	& $^{205}$Pb & -0.148(5) & - & 0.13(4) \\
\hline
\end{tabular}
\end{center}
\end{table*}

\begin{table*}[!htb]
\caption{\label{tab:expdata2} The experimental data for the $B(E2)$ trends of ${2}^+$ states in W.U. for even-even Hg, Pb and Po isotopes. Experimental data have been taken from the evaluated $B(E2)$ compilation of Pritychenko $et$ $al.$ \cite{pritychenko2016}, unless otherwise stated. The uncertainties are shown in the parentheses. }
\begin{center}
\begin{tabular}{c c c c}
\hline
N & Hg isotopes & Pb isotopes & Po isotopes  \\
\hline
&\multicolumn{3}{c}{$B(E2)$ W.U. ~\cite{pritychenko2016}}\\
\hline
100 &  48(7) &  - & - \\	
102	&  54.8(29)	&  - & - \\
104	&  52.2(25)	&  6.0(17) & - \\
106 & 46.6(66) &  8.0(27) & - \\
108	&  53.8(84) &  - & - \\
110	&  45(3)~\cite{ensdf} &  - &  90($^{+20}_{-14}$)\\
112	&  42($^{+26}_{-12}$)~\cite{xundl} &  - &  47.0(62) \\
114	&  39($^{+9}_{-6}$)~\cite{xundl} &  18.2($^{+48}_{-41}$)~\cite{xundl} &  37.9($^{+85}_{-70}$) \\
116	&  33.8(24) &  13.1($^{+49}_{-35}$)~\cite{xundl} &  30.5(17) \\
118	&  28.05(20) &  - & 31.8($^{+97}_{-74}$) \\
120	&  24.6(81)	&  $>$ 0.0975	& - \\
122	&  17.47(60) &  4.45(19) &  18($^{+14}_{-10}$)~\cite{xundl} \\
124	&  11.89(59) &  2.737(77) & - \\
126 &  - &  7.84(49) &  0.54(11) \\
\hline
\end{tabular}
\end{center}
\end{table*}

\begin{centering}
\begin{figure}[!htb]
\includegraphics[width=14cm,height=12cm]{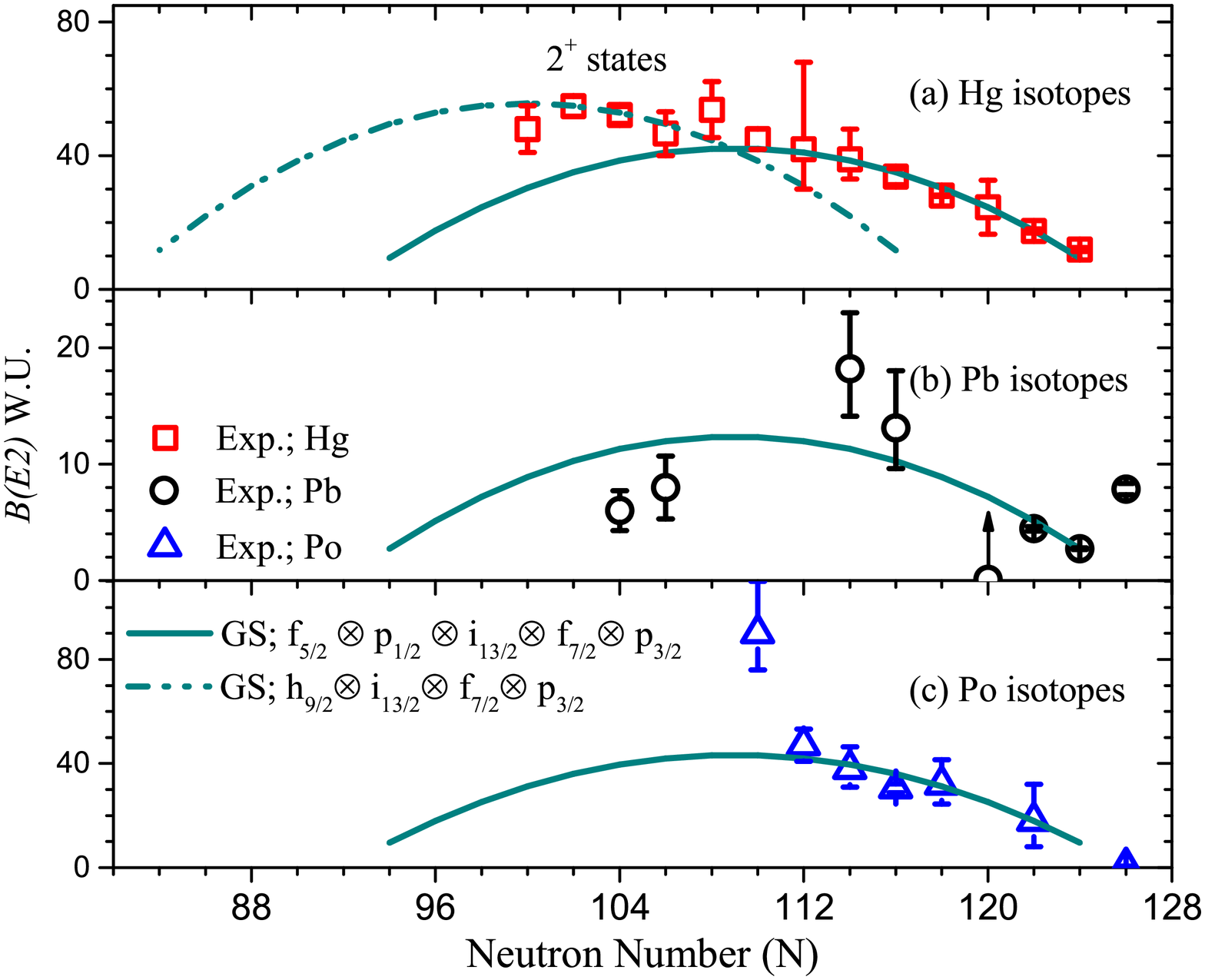}
\caption{\label{fig:be2_2}(Color online) Experimental \cite{pritychenko2016} and calculated $B(E2)$ trends for the first $2^+$ states in Hg, Pb and Po isotopes. To complete the systematics of experimental values, the data have also been adopted from \cite{ensdf,xundl}; refer text for details. The GS calculations for heavier isotopes are done by using $\Omega=17$ corresponding to \{$\tilde{j}=i_{13/2} \otimes f_{7/2} \otimes p_{3/2} \otimes f_{5/2} \otimes p_{1/2}$\} configuration. For lighter Hg isotopes, GS calculated results using $\Omega=18$  corresponding to \{$\tilde{j}=h_{9/2} \otimes i_{13/2} \otimes f_{7/2} \otimes p_{3/2} \}$ configuration are also shown.} 
\end{figure}
\end{centering}

\section{$B(E2)$ trends for the first $2^+$ states in Hg, Pb and Po isotopes}

We also discuss $B(E2)$ trends for the first $2^+$ states in even-A Hg, Pb and Po isotopes. It also supports a double-hump behavior in all the three Hg, Pb and Po isotopic chains, similar to the analogy discussed earlier in Cd, Sn and Te isotopic chains~\cite{maheshwari20191,maheshwari2020}. The measured $B(E2)$ values have been taken from the compilation made by Pritychenko $et$ $al.$ in 2016~\cite{pritychenko2016}. To complete the systematics, we have adopted values from ENSDF ~\cite{ensdf} and XUNDL (Experimental Unevaluated Nuclear Data List Search and Retrieval) ~\cite{xundl} data bases; refer Table~\ref{tab:expdata2} for the values.

Fig. \ref{fig:be2_2}(a) presents a comparison of experimental~\cite{pritychenko2016} and GS calculated $B(E2)$ trends for the first $2^+$ states in Hg isotopes. For $N=110$, $^{190}$Hg, the measured $B(E2)$ value has been adopted from XUNDL data base~\cite{xundl}, while for $^{192,194}$Hg ($N=112,114$), the ENSDF~\cite{ensdf} values have been adopted; as mentioned in Table~\ref{tab:expdata2}.  The GS calculations have been carried out using the $\Omega=18$ and $\Omega=17$, before and after the middle of the valence space $(N=104)$, respectively, as shown in Fig. \ref{fig:be2_2}(a). A single-point fitting to the experimental data has been used to take care of the structural effects. The $\Omega=18$ refers to the multi-j configuration of \{$\tilde{j}=h_{9/2} \otimes i_{13/2} \otimes f_{7/2} \otimes p_{3/2} $\} with a natural choice of neutron core at $N=82$, while the $\Omega=17$ refers to the multi-j configuration of \{$\tilde{j}=i_{13/2} \otimes f_{7/2} \otimes p_{3/2} \otimes f_{5/2} \otimes p_{1/2}$\} by freezing $h_{9/2}$ orbital resulting in a core at $N=92$. The GS calculated trend explains the experimental data quite well, supporting a change in the orbitals around middle of the valence space i.e. $N=104$. This is very similar to the previous interpretation of double-hump parabolic behavior of $B(E2)$ values for the first $2^+$ states in Cd, Sn and Te isotopes~\cite{maheshwari20191}. 

Similarly, we have compared the experimental~\cite{pritychenko2016} and GS calculated $B(E2)$ trends for the first $2^+$ states in Pb and Po isotopes, respectively, in Figs. \ref{fig:be2_2}(b) and \ref{fig:be2_2}(c). Only a lower limit as $> 0.0975$ W.U. is known experimentally for $N=120$, $^{202}$Pb~\cite{pritychenko2016}. For $^{196,198}$Pb ($N=114,116$) and $^{206}$Po ($N=122$), the experimental data have been taken from ENSDF data base~\cite{ensdf}. Table~\ref{tab:expdata2} lists all the $B(E2)$ values adopted in Fig~\ref{fig:be2_2}. The calculated trend from $\Omega=17$ has been shown for comparison. Very limited data are available in Pb isotopes making it difficult to conclude the situation. Further measurements are hence needed. The GS calculated trend using $\Omega=17$ explains the measured values for Po isotopes with $N \ge 110$ quite well. The new measurements towards lighter Po isotopes ($N<108$) are required to complete the picture. 

\section{Predictions and future outlook}

Using GSSM, the $g-$factor estimates have been made for several nuclei: (i) ${13/2}^+$ isomers in odd-A $^{173-185}$Hg, $^{181-183}$Pb, and $^{181-193}$Po isotopes; (ii) the ${12}^+$ isomers in even-A $^{174-194}$Hg, $^{176-190}$Pb, and $^{178-196}$Po isotopes; (iii) ${33/2}^+$ isomers in odd-A $^{173-197}$Hg, $^{175-191,195-199}$Pb, and $^{179-201}$Po isotopes. The multi-j \{$\tilde{j}=i_{13/2} \otimes f_{7/2} \otimes p_{3/2} $\} configuration remains dominant for the $v=1$, ${13/2}^+$; $v=2$, ${12}^+$; and $v=3$, ${33/2}^+$ isomers in all the three Hg, Pb and Po isotopic chains. The GSSM line corresponding to this configuration lies at $-0.153$ $n.m.$ value. Similarly, one can predict the corresponding $Q-$moment and $B(E2)$ trends for these states, as discussed in the previous sections. With the present experimental facilities in nuclear physics, most of the nuclei can now be accessed. New measurements are hence needed to confirm the predictions based on the systematic trends.

\section{Conclusion}

In this paper, we have studied the ${13/2}^+$, ${12}^+$, and ${33/2}^+$ isomers in all the three Hg$(Z=80)$, Pb$(Z=82)$ and Po$(Z=84)$ isotopic chains using generalized seniority scheme. The complex systems of Hg, Pb and Po isotopes display regular patterns of nuclear moments, particularly the $g-$ factors and $Q-$moments. A consistently same multi-j configuration is able to explain the various nuclear properties including transition probabilities and moments. The reason for such simple trends is the presence of many-body symmetries resulting in a collective nucleonic motion. These ${13/2}^+$, ${12}^+$, and ${33/2}^+$ isomers are described as the generalized seniority $v=1$, $v=2$ and $v=3$ isomers, respectively, from the same configuration. In addition to this, we have studied the first $2^+$ states in Hg, Pb and Po isotopes and explained the origin of inverted $B(E2)$ parabolic trends using generalized seniority scheme. The success of GSSM in explaining the $g-$factor trends in and around Pb isotopes is quite encouraging. Also, the GS scheme describes the $Q-$moment and $B(E2)$ trends for these generalized seniority isomers quite well. Predictions have been made for several nuclei. Dedicated experimental studies would be required to refine the systematics in Hg, Pb and Po isotopes.

\section*{Acknowledgements}

BM and AKJ thank Amity University for providing the support and facilities to carry out this work. One of us (AKJ) acknowledges the financial support received from S.E.R.B. (Govt. of India) in the form of a research grant. DC acknowledges the support and facilities received from IIT Ropar to complete this work.



\begin{thebibliography}{150}

\bibitem{bohr1975} A. Bohr and B. R. Mottelson, Nuclear Structure, W. A. Benjamin Inc., New York, Volume I: Single Particle Motion, 1969 and Volume II: Nuclear Deformations, 1975.
\bibitem{mayer1955} M. G. Mayer and J. H. D. Jensen, Elementary Theory of Nuclear Shell Structure, John Wiley and Sons Inc., New York, 1955.
\bibitem{frank2009} A. Frank, J. Jolie and P. Van Isacker, Symmetries in atomic nuclei, Springer Tracks in Modern Physics, $\textbf{230}$, Springer 2009.
\bibitem{casten1990} R. F. Casten, Nuclear Structure from a Simple Perspective, Oxford University Press, 2000.
\bibitem{jain2021} A. K. Jain, B. Maheshwari, A. Goel, Nuclear Isomers- A Primer, Springer Nature, 2021. 
\bibitem{talmi1993} I. Talmi, Simple Models of Complex Nuclei, Harwood Academy (1993).
\bibitem{racah1943} G. Racah, Phys. Rev $\textbf{63}$, 367 (1943).
\bibitem{racah1952} G. Racah, Research Council of Israel, Jerusalem, $\textbf{L. Farkas Memorial Volume}$, 294 (1952).
\bibitem{shalit1963} A. de Shalit and I. Talmi, Nuclear Shell Theory, Academic Press, New York and London, 1963.
\bibitem{heyde1990} K. Heyde, The Nuclear Shell model, Springer-Verlag Belrin Heidelberg, 1990.
\bibitem{isacker2014} P. V. Isacker, Nucl. Phys. News $\textbf{24}$, 23 (2014).
\bibitem{bengtsson1989} R. Bengtsson and W. Nazarewicz, Zeitschrift für Physik A Atomic Nuclei volume $\textbf{334}$, 269 (1989).
\bibitem{frank2004} A. Frank, P. Van Isacker, and Carlos E. Vargas, Phys. Rev. C $\textbf{69}$, 034323 (2004).
\bibitem{heyde2011} K. Heyde and J. Wood, Rewiews of Modern Physics, $\textbf{83}$ (2011).
\bibitem{maheshwari2016} B. Maheshwari and A. K. Jain, Phys. Lett. B $\textbf{753}$, 122 (2016).
\bibitem{maheshwari20161} B. Maheshwari, A. K. Jain and B. Singh, Nucl. Phys. A $\textbf{952}$, 62 (2016).
\bibitem{jain2017} A. K. Jain and B. Maheshwari, Nucl. Phys. Rev. $\textbf{34}$, 73 (2017). 
\bibitem{jain20171} A. K. Jain and B. Maheshwari, Physica Scripta $\textbf{92}$, 074004 (2017).
\bibitem{maheshwari2019} B. Maheshwari and A. K. Jain, Nucl. Phys. A $\textbf{986}$, 232 (2019).
\bibitem{maheshwari20191} B. Maheshwari, H. A. Kassim, N. Yusof and A. K. Jain, Nuclear Physics A $\textbf{992}$, 121619 (2019).
\bibitem{neyens2003} G. Neyens, Rep. Prog. Phys. $\textbf{66}$, 633 (2003).
\bibitem{flowers1952} B.H. Flowers, Proc. Roy. Soc. (London) A, $\textbf{212}$, 248 (1952).
\bibitem{kerman1961} A. K. Kerman, Ann. Phys. (NY) $\textbf{12}$ 300 (1961).
\bibitem{helmers1961} K. Helmers, Nucl. Phys. $\textbf{23}$, 594 (1961).
\bibitem{arima1966} A. Arima and M. Ichimura, Prog. of Theo. Phys. $\textbf{36}$, 296 (1966).
\bibitem{talmi1971} I. Talmi, Nucl. Phys. A $\textbf{172}$, 1 (1971).
\bibitem{shlomo1972} S. Shlomo and I. Talmi, Nucl. Phys. A $\textbf{198}$, 82 (1972).
\bibitem{arvieu1966} R. Arvieu and S.A. Moszokowski, Phys. Rev. $\textbf{145}$, 830 (1966).
\bibitem{maheshwari2017} B. Maheshwari, S. Garg and A. K. Jain, Pramana-Journal of Physics (Rapid Communication) $\textbf{89}$, 75 (2017).
\bibitem{maheshwarijnp}B. Maheshwari, J. Nucl. Phy. Mat. Sci. Rad. A. $\textbf{6}$, 147 (2019).
\bibitem{maheshwari2020} B. Maheshwari, Euro. Phys. Jour. Special Topics $\textbf{229}$, 2485 (2020). 
\bibitem{maheshwari20201} B. Maheshwari and B. K. Agrawal, Nuclear Structure Physics $\textbf{231}$, CRC Press (2020).
\bibitem{agrawal2020} B. K. Agrawal and B. Maheshwari, Euro. Phys. Jour. Special Topics $\textbf{229}$, 2459 (2020).
\bibitem{kota2017} V. K. B. Kota, Bulgarian Journal of Physics $\textbf{44}$, 454 (2017); arXiv: 1707.03552v1.
\bibitem{schmidt1937} T. Schmidt, Z. Physik $\textbf{106}$, 358 (1937).
\bibitem{nomura1972} M. Nomura, Phys. Lett. B $\textbf{40}$, 522 (1972).
\bibitem{zamick1971} L. Zamick, Phys. Lett. B $\textbf{34}$, 472 (1971).
\bibitem{towner1987} I. S. Towner, Phys. Reports $\textbf{155}$, 263 (1987).
\bibitem{ensdf} ENSDF data base; www.nndc.bnl.gov/ensdf/.
\bibitem{stone2019} N. J. Stone, www-nds.iaea.org/publications, INDC(NDS)-0794, Nov. (2019).
\bibitem{stone2014} N. J. Stone, www-nds.iaea.org/publications, INDC(NDS)-0658, Feb. (2014).
\bibitem{wouters1991} J. Wouters, N. Severijns, J. Vanhaverbeke and L. Vanneste, J. Phys. G $\textbf{17}$, 1673 (1991).
\bibitem{mahnke1979} H.-E. Mahnke, $et$ $al.$, Phys. Lett, $\textbf{88B}$, 48 (1979).
\bibitem{dracoulis1998} G. D. Dracoulis, A. P. Byrne, A. M. Baxter, Phys. Lett. B $\textbf{432}$, 37 (1998).
\bibitem{bricc}BRICC: http://bricc.anu.edu.au/
\bibitem{wrzesinski2015} J. Wrzesinski $et$ $al.$, Phys. Rev. C $\textbf{92}$, 044327 (2015).
\bibitem{pritychenko2016} B. Pritychenko, M. Birch, B. Singh and M. Horoi, Atomic Data Nuclear Data Tables $\textbf{107}$, 1 (2016).
\bibitem{xundl} XUNDL data base: https://www.nndc.bnl.gov/ensdf/ensdf/xundl.jsp

\end{thebibliography}
\end{document}